\documentclass{article}
\usepackage{spconf,amsmath,graphicx}
\usepackage{epstopdf}
\usepackage{lipsum}
\usepackage{todonotes}
\usepackage{textcomp}
\usepackage{xspace}
\usepackage{color}
\usepackage[absolute]{textpos}
\usepackage[bottom]{footmisc}
\usetikzlibrary{spy}
\usepackage{subcaption}
\usepackage{xspace}
\usepackage{paralist}
\usepackage{multirow}
\usepackage{hyperref}

\usepackage{tcolorbox}
\usepackage{tabularx}
\usepackage{array}
\usepackage{colortbl}
\tcbuselibrary{skins}
\usepackage{diagbox}
\usepackage{float}
\usepackage{arydshln}
\usepackage{nccmath} 
\usepackage[abs]{overpic} 

\newcommand{\sign}{\boldsymbol{s}}

\newcommand{\img}{\boldsymbol{x}}

\newcommand{\alfk}{\alpha_k(s_i)}
\newcommand{\alfkp}{\alpha_{k'}(s_i)}

\newcommand{\polyAlf}{g^\alpha_k(s_i)}
\newcommand{\polyAlfP}{g^\alpha_{k'}(s_i)}
\newcommand{\polyMu}{g^\mu_k(s_i)}
\newcommand{\polyMuP}{g^\mu_{k'}(s_i)}
\newcommand{\polyVar}{g^{\sigma}_k(s_i)}

\newcommand{\muk}{\mu_k(s_i)}
\newcommand{\vark}{\sigma^{2}_k(s_i)}

\DeclareMathOperator*{\argmax}{arg\,max}

\tikzset{annot/.style={draw=black,fill=white,text=black}}

\makeatletter
\DeclareRobustCommand\onedot{\futurelet\@let@token\@onedot}
\def\@onedot{\ifx\@let@token.\else.\null\fi\xspace}
\def\eg{\emph{e.g}\onedot}

\makeatother

\title{Fully Unsupervised Probabilistic Noise2Void}
\name{
Mangal Prakash$^{1,2\ast}$, Manan Lalit$^{1,2\ast}$, Pavel Tomancak$^{2}$,}
\secondlinename{Alexander Krull$^{1,2\dagger}$, Florian Jug$^{1,2\dagger}$}
\address{
$^{1}$Center for Systems Biology Dresden (CSBD)\\
$^{2}$Max-Planck Institute of Molecular Cell Biology and Genetics\\
$^{\ast}$equal contribution \, \, 
$^{\dagger}$joint supervision
}

\newcommand\figTeaser{
\begin{figure}[h]
\centering
\begin{overpic}
  [width=0.23\textwidth,trim=0 8 0 0,clip=true]
  {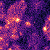}\hspace{0.1em}
  \put(5,64){\color{white}\textbf{RAW}}
\end{overpic}
\begin{overpic}
  [width=0.23\textwidth,trim=0 8 0 0,clip=true]
  {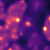}\hspace{0.1em}
  \put(5,64){\color{white}\textbf{Ours}}
\end{overpic} \\
\vspace{0.1em}
\begin{overpic}
  [width=0.23\textwidth,trim=0 8 0 0,clip=true]
  {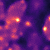}\hspace{0.1em}
  \put(5,64){\color{white}\textbf{Noise2Void}}
\end{overpic}
\begin{overpic}
  [width=0.23\textwidth,trim=0 8 0 0,clip=true]
  {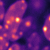}\hspace{0.1em}
  \put(5,64){\color{white}\textbf{Ground Truth}}
\end{overpic}
\caption{\vspace{-0.2em}
Our proposed GMM bootstrapping approach does not require paired training or calibration data, but achieves superior results compared to other fully unsupervised methods.}
  \label{fig:Teaser}
\end{figure}
}

\newcommand\figQualitative{
\begin{figure*}[hbt]
\centerline{
\includegraphics[trim=0pt 410pt 0pt 243pt, scale=0.4, clip=true,width=\linewidth]{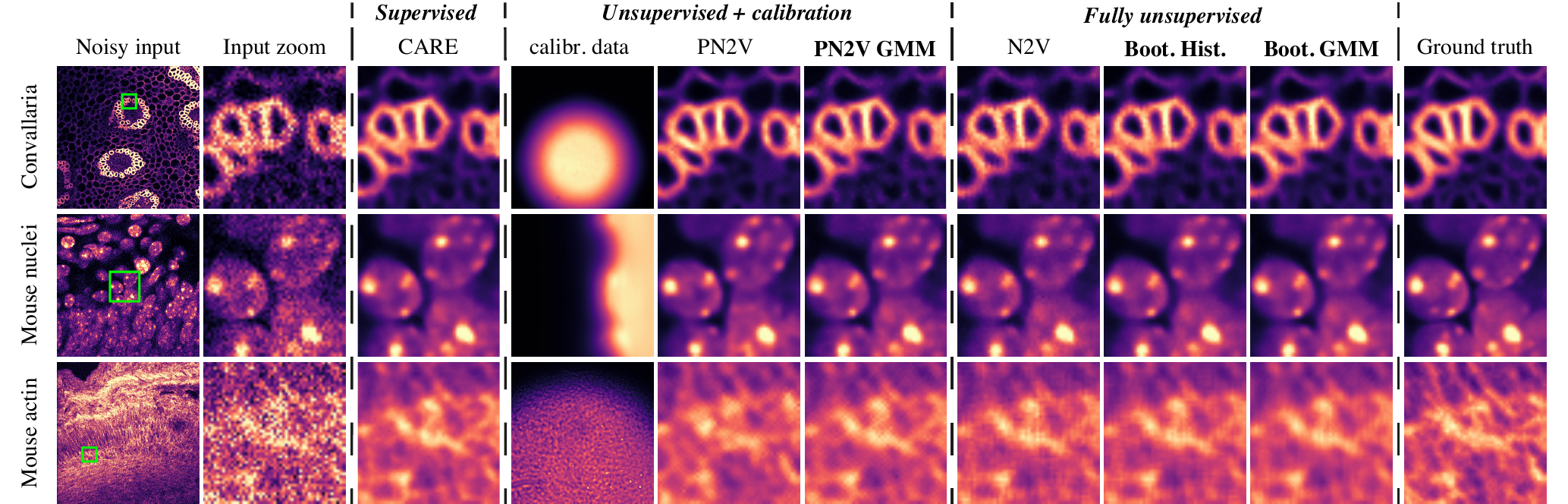}
}
\caption{A visual comparison of results obtained by CARE, N2V, PN2V, and our proposed methods (bold).
We distinguish three families of methods: fully supervised (CARE), unsupervised but requiring additional calibration data (PN2V, our PN2V GMM), and fully unsupervised (N2V, PN2V using our bootstrapped histogram and GMM based noise models).
The leftmost column in the unsupervised+calibration category shows the average of all available calibration images used for PN2V and PN2V GMM (see main text).
Note that results of our fully unsupervised methods reach very similar quality to methods requiring either clean GT, or additional calibration data.
}
\vspace{-3mm}
\label{fig:QualitativeResults}
\end{figure*}
}

\newcommand\tabResults{
\begin{table}[tb]
\begin{center}
 \begin{tabular}{ p{1.8cm} p{1.7cm} p{1.7cm} p{1.7cm}}
    \hline
  {\small Methods} & {\small \textit{Convallaria}} & {\small Mouse nuclei} & {\small Mouse actin} \\
 \hline\hline
 {\small CARE} & \small{\textbf{36.71$\pm$0.026}} & \small{\textbf{36.58$\pm$0.019}}& \small{\textbf{34.20$\pm$0.021}} \\ 
 \hdashline
 {\small PN2V} & \small{\textbf{36.51$\pm$0.025}} & \small{36.29$\pm$0.007} & \small{33.78$\pm$0.006} \\

 \textbf{{\small PN2V GMM}} & \small{36.47$\pm$0.031} & \small{\textbf{36.35$\pm$0.018}} & \small{\textbf{33.86$\pm$0.018}} \\
 \hdashline
 {\small N2V} & \small{35.73$\pm$0.037} & \small{35.84$\pm$0.015} & \small{33.39$\pm$0.014} \\

 \textbf{{\small Boot. Hist.}} & \small{36.19$\pm$0.016} & \small{36.31$\pm$0.013} & \small{33.61$\pm$0.016}\\

 \textbf{{\small Boot. GMM}} & \small{\textbf{36.70$\pm$0.012}} & \small{\textbf{36.43$\pm$0.014}} & \small{\textbf{33.74$\pm$0.012}} \\ 
\hline
\end{tabular}

\captionof{table}{Comparision of the denoising performance of all tested methods.
Mean PSNR and $\pm 1$ standard error over five repetitions of each experiment are shown.
Names of our proposed methods are shown in bold.
Bold numbers indicate the best performing method in its respective category (supervised, unsupervised + calibration, and fully unsupervised; from top to bottom, separated by dashed lines). 
}
\label{QuantitativeTable}
\vspace{-8mm}
\end{center}
\end{table}
}
\newcommand\tabBestParameters{
\begin{table}[bt]
\begin{center}
\begin{tabular}{ccc}\hline
Gaussians & Two coefficients & Three coefficients \\ \hline\hline
  \small{1} & \small{36.56$\pm$0.022} & \small{36.34$\pm$0.040} \\ 
  \small{2} & \small{36.48$\pm$0.020} & \small{36.35$\pm$0.014} \\ 
  \small{3} & \small{36.47$\pm$0.031} & \small{36.31$\pm$0.022} \\ \hline
\end{tabular}
\captionof{table}{Testing a variety of GMM hyper-parameters.
We tested GMMs using one, two, and three Gaussians, each using linear ($n=2$) and quadratic ($n=3$) parametrizations (see Section~\ref{sec:methods}), to denoise the \textit{Convallaria} data. 
The table shows the mean PSNR and standard error over 5 repetitions for each setup.}
\label{AblationTable}
\vspace{-5mm}
\end{center}

\end{table}
}

\newcommand\tabgaussianComparison{
\begin{table}[h]

\begin{center}
\begin{tabular}{cccccc}\hline
Gaussians & NM1 & NM2 & NM3 & NM4 & NM5 \\ \hline \hline
\small{$1$} & \small{$36.56 $} & \small{$36.03 $} & \small{$35.98$} & \small{$35.85 $} & \small{$35.78$} \\
\small{$3$} & \small{$36.47 $} & \small{$36.58 $} & \small{$36.37 $} & \small{$36.20 $} & \small{$36.08 $} \\ \hline
\end{tabular}
\captionof{table}{Denoising performance of PN2V GMM with linear noise models using one versus three Gaussians. For each case, five noise models were derived from different subsets of the available calibration data (see Fig.~\ref{fig:SystematicTests}). We report the mean PSNR over 5 repetitions for each setup.}
\label{gaussiancomparison}
\vspace{-5mm}
\end{center}
\end{table}
}
\newcommand\figSystematicTests{
\begin{figure}[t]

\centering
\begin{overpic}[width=0.24\textwidth]
  {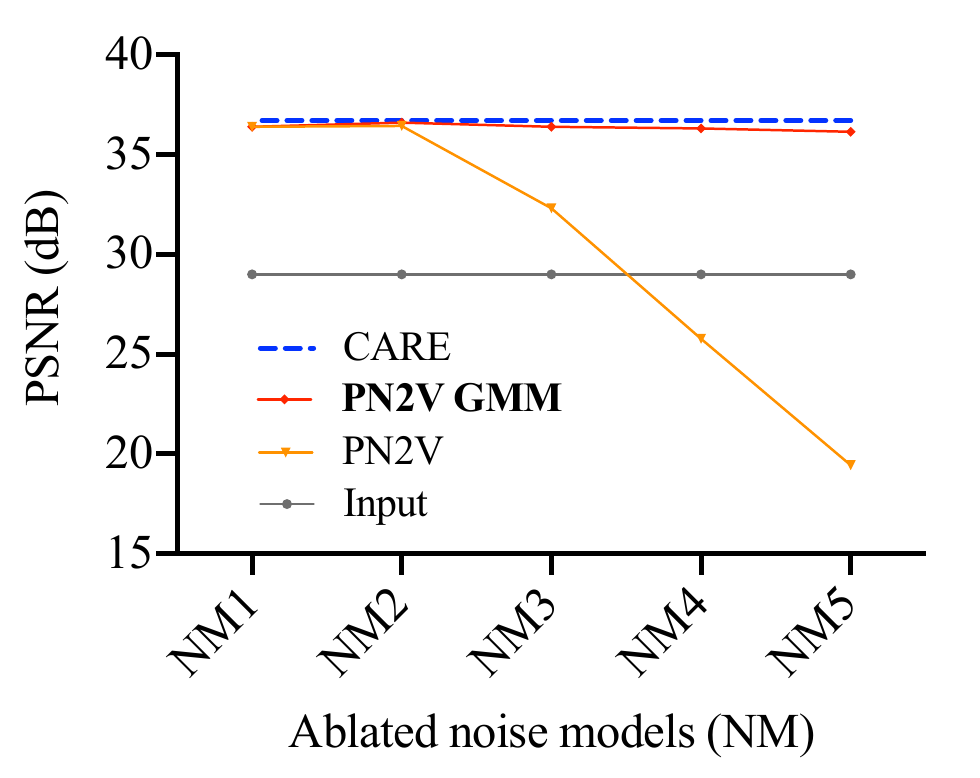}
 \put(-1, 85){\textbf{(a)}}  
\end{overpic}
\begin{overpic}[width=0.24\textwidth]
  {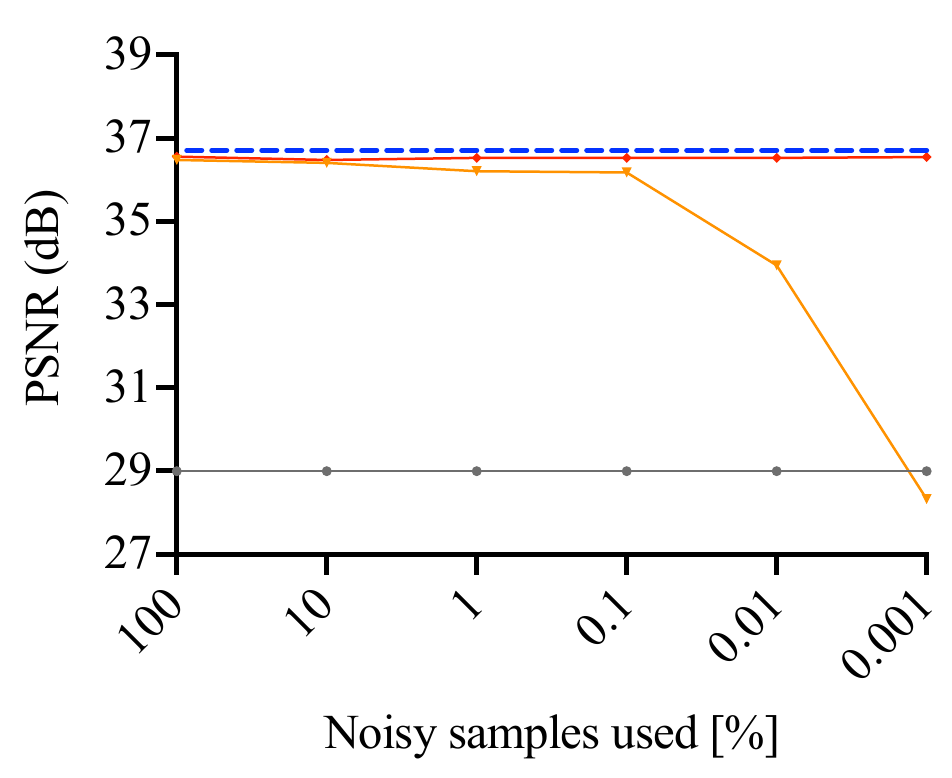} 
  \put(-1, 85){\textbf{(b)}}
\end{overpic}

\caption{Ablation studies on \textit{Convallaria} data.
Denoising performance of PN2V with histogram and linear GMM noise models is shown. 
{\bf (a)} The five noise models we tested are deduced from subsets of the available calibration data, such that: the entire range of signals used in the \textit{Convallaria} data is covered (NM1), only the lower 40\% are covered (NM2), the lower 25\% (NM3), 15\% (NM4) and 9\% (NM5).
{\bf (b)} This case is obtained by reducing the fraction of available noisy calibration pixels from NM1, via random subsampling.
}
  \label{fig:SystematicTests}
\end{figure}
}

\newcommand\figAblation{
\begin{figure}[hbt]
\centering
\begin{overpic}[width=0.48\textwidth]
  {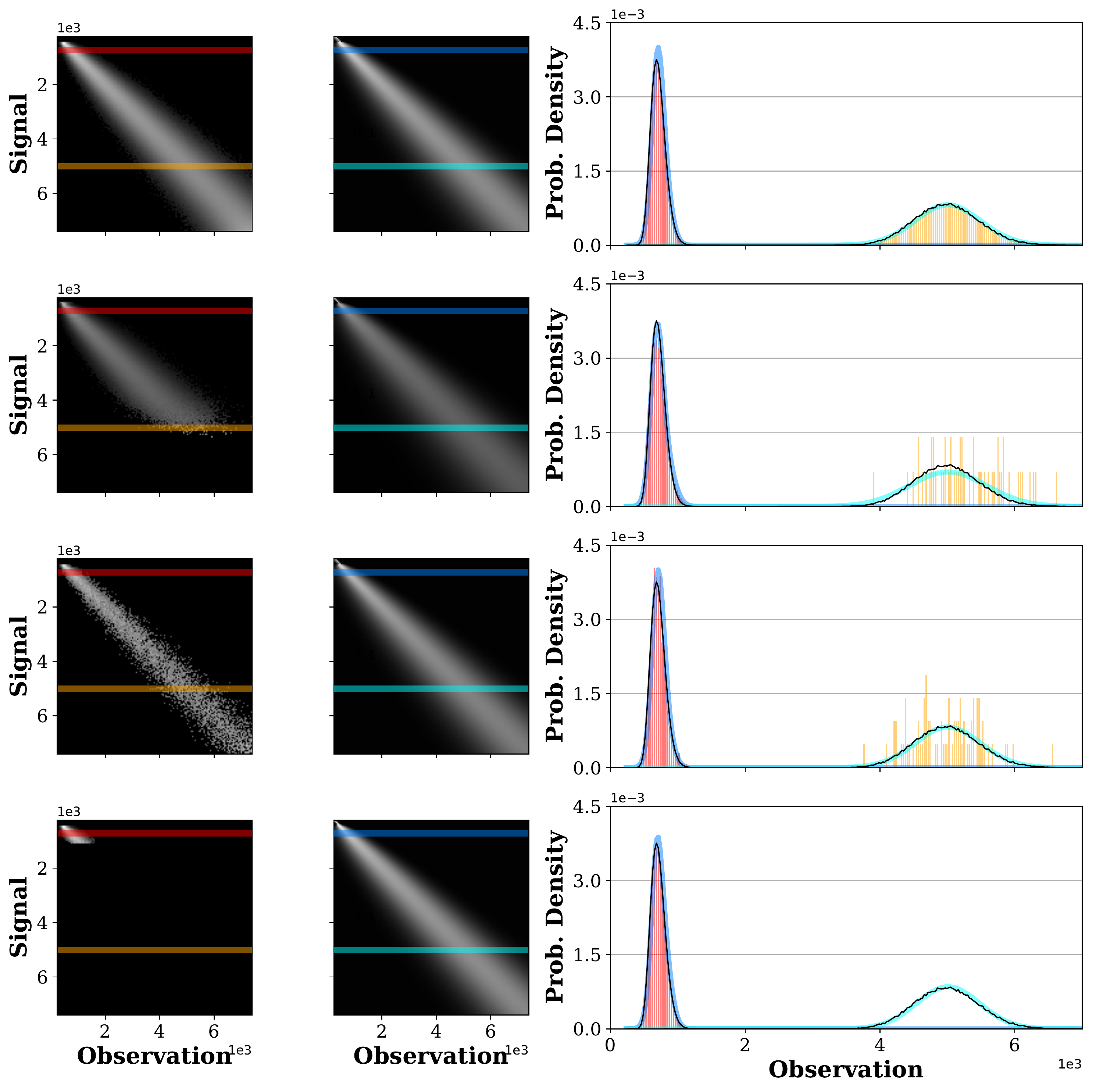}
\put(-1, 230){\color{black}\textbf{(a)}}  
\put(-1, 172){\color{black}\textbf{(b)}} 
\put(-1, 114){\color{black}\textbf{(c)}} 
\put(-1, 56){\color{black}\textbf{(d)}} 
\end{overpic}
\caption{
Noise models for \emph{Convallaria} data.
Left column shows histogram based noise models, the center column their respective GMM based equivalent. 
Rightmost column shows noise models for specific signals (colors, histograms shown as vertical lines).
For comparison, full calibration data histogram is always included as black curve. 
{\bf (a)} Noise models trained on full calibration data.
{\bf (b)} Bootstrapped noise models.
{\bf (c)} Noise models trained on sub-sampled ($0.1\%$) calibration data~(Fig.~\ref{fig:SystematicTests}b).
{\bf (d)} Noise models trained on reduced available calibration data~(NM5 from Fig.~\ref{fig:SystematicTests}a).
}
\vspace{-3mm}
\label{fig:NoiseModelViews}
\end{figure}
}

\begin{document}
%\ninept
%
\maketitle
\begin{abstract}
Image denoising is the first step in many biomedical image analysis pipelines and Deep Learning (DL) based methods are currently best performing.
A new category of DL methods such as Noise2Void or Noise2Self can be used fully unsupervised, requiring nothing but the noisy data.
However, this comes at the price of reduced reconstruction quality.
The recently proposed Probabilistic Noise2Void (PN2V) improves results, but requires an additional noise model for which calibration data needs to be acquired.
Here, we present improvements to PN2V that 
$(i)$~replace histogram based noise models by parametric noise models, and 
$(ii)$~show how suitable noise models can be created even in the absence of calibration data. This is a major step since it actually renders PN2V fully unsupervised.
We demonstrate that all proposed improvements are not only \mbox{academic} but indeed relevant.
\end{abstract}
%
% %%%%%%%%%%%%%%%%%%%%%%%%%%%%%%%%%%%%%%%%%%%%%%%%%%%%%%%%%%%%%%%%%%%%%
\section{Introduction}
\label{sec:intro}
\vspace{-2mm}
% %%%%%%%%%%%%%%%%%%%%%%%%%%%%%%%%%%%%%%%%%%%%%%%%%%%%%%%%%%%%%%%%%%%%%
With the advent of Deep Learning (DL), the field of biomedical image denoising has recently taken rapid strides~\cite{weigert2018content,zhang2019poisson, buchholz2019cryo,buchholz2019content,krull2019noise2void,krull2019probabilistic, belthangady2019applications}.
Today, Content-Aware image REstoration (CARE) methods are leading the field due to their \textit{content awareness} -- learning a strong prior on the visual nature of the data to be reconstructed \cite{weigert2018content,sui2018differential,laine2019nanoj,ouyang2018deep}. 

\sloppy 
\figTeaser

While CARE was initially proposed using pairs of noisy and clean (ground truth) images during training, several ways to circumvent this requirement have been proposed.
Noise2Noise~\cite{noise2noise} shows how corresponding noisy image pairs can lead to virtually the same results.
Self-supervised models, like Noise2Void~(N2V)~\cite{krull2019noise2void} and Noise2Self~\cite{batson2019noise2self} show how even the requirement for a second noisy image can be avoided.
These methods can train directly on the body of data to be denoised, making them extremely useful for practical applications.
However, self-supervised methods are known to perform less well than models trained using paired training data~\cite{krull2019noise2void,krull2019probabilistic}. 

The recently proposed Probabilistic Noise2Void (PN2V)~\cite{krull2019probabilistic} shows how using sensor specific noise models can improve the quality of self-supervised denoising, bringing it close to traditional paired training. A PN2V noise model is computed from a sequence of noisy calibration images and characterizes the distribution of noisy pixel around their respective ground truth signal value.
In the context of PN2V, noise models are a collection of histograms~\cite{krull2019probabilistic}.

In this work we make three major contributions.
$(i)$~We improve PN2V by introducing parametric noise models based on Gaussian Mixture Models (GMM) and show why they perform better than histogram based representations.
$(ii)$~We show how to bootstrap a suitable noise model, even in the absence of calibration data. This renders PN2V fully unsupervised, where nothing besides the data to be denoised is required for the method to be applied. 
$(iii)$~All calibration data and corresponding noisy image data is made publicly available together with the code ({\small \url{github.com/juglab/ppn2v}}).

\figQualitative

\tabResults
% %%%%%%%%%%%%%%%%%%%%%%%%%%%%%%%%%%%%%%%%%%%%%%%%%%%%%%%%%%%%%%%%%%%%%
\section{Proposed Approaches and Methods}
\label{sec:methods}
\vspace{-2mm}
% %%%%%%%%%%%%%%%%%%%%%%%%%%%%%%%%%%%%%%%%%%%%%%%%%%%%%%%%%%%%%%%%%%%%%

\sloppy

\noindent \textbf{Histogram based noise models},  
% - - - - - - - - - - - - - - - - - - - - - - - - - - - - -
as originally suggested for PN2V, are built from a stack of calibration images $\img^1 ,\dots , \img^m$.
The imaged structures in this sequence can be arbitrary but must be static.
Such images can, for example, be recorded by imaging the back illuminated half opened field diaphragm (see Fig.~\ref{fig:QualitativeResults}). 
In order to minimize the effects of vibrations and sample drift, we recommend to acquire calibration data in defocus.
We call the average signal $\sign = \frac{1}{m} \sum_{j=1}^m \img^j$ ground truth (GT). 
It is an established protocol to average multiple static but noisy acquisitions to obtain a corresponding GT image~\cite{zhang2019poisson}.
By discretizing each GT pixel signal $s_i$ and corresponding noisy observations $x^j_i$, a histogram can be created for each GT signal covering all corresponding noisy observations.
The normalized set of histograms constitutes the camera noise model  used in PN2V~\cite{krull2019probabilistic}, describing the distribution of noisy pixel values $p(x_i|s_i)$ that are to be expected for each GT signal.

\fussy

\vspace{1mm}
\noindent \textbf{GMM based noise models}
% - - - - - - - - - - - - - - - - - - - - - - - - - - - - -
describe the distribution of noisy observations $x_i$ for a GT signal $s_i$ as the weighted average of $K$ normal distributions: \useshortskip
\begin{equation}
    p(x_i|s_i) = \sum_{k=1}^{K}\alfk f\big( \muk,\vark \big),
    \label{eq:GMM}
\end{equation}
where $f\big( \muk,\vark \big)$ is the probability density function of the the normal distribution. 
We define each component's weight $\alfk$, mean $\muk$, and variance $\vark$ as a function of the signal $s_i$. To ensure all weights are positive and sum to one we define \useshortskip
\begin{equation}
\alfk= \exp \big(\polyAlf \big) / \sum_{k'=1}^K \exp \big( \polyAlfP \big),
\end{equation}
where $\polyAlfP$ is a polynomial of degree $n$.
To ensure that our distributions are always centered around the true signal $s_i$, we  define %\useshortskip
%\begin{equation}
$\muk= s_i + \polyMu - \sum_{k'=1}^K  \alfkp \polyMuP$,
%\end{equation}
where $\polyMu$ is again a polynomial of degree $n$.
Finally, to ensure numerical stability, we define the variance %\useshortskip
%\begin{equation}
$\vark = \max(\polyVar,c)$,
%\end{equation}
where $c=50$ is a constant, and $\polyVar$ is again a polynomial of degree $n$. Hence, our GMM based noise model is fully described by the $3 \times K \times n$ long vector of polynomial coefficients $\boldsymbol{a}$.
We use a maximum likelihood approach to fit the parameters to our calibration data, optimizing for \useshortskip
\begin{equation}
\argmax_{\boldsymbol{a}} \sum_{i,j} 
\log{p(
x_i^j| s_i 
)},
\label{eq:opt}
\end{equation}
where $p(x_i^j| s_i)$, is the GMM as described in Eq.~\ref{eq:GMM}.
We use numerical optimization, see Section \ref{sec:results}.

\figSystematicTests

\tabBestParameters

\tabgaussianComparison

\label{sssec:bootstrap}

\vspace{1mm}
\noindent \textbf{Bootstrapped PN2V} allows us to address the scenarios where no calibration data is available, \eg, data that was acquired without denoising in mind.
We propose the following bootstrapping procedure. 
First, we train and apply the unsupervised N2V~\cite{krull2019noise2void} on the body of available noisy images $\img^j$. 
Then, we treat the resulting denoised images $\hat{\sign}^j$ as if they were the GT, henceforth calling them pseudo ground truth. 
We can now use the corresponding noisy $x_i^j$ and denoised $\hat{s}_i^j$ pixel values to either construct a histogram or learn a GMM based noise model. 

\section{Experiments and Results}
\label{sec:results}
\vspace{-2mm}

\noindent \textbf{Datasets:} 
We acquired three datasets (Fig.~\ref{fig:QualitativeResults}) which are made publicly available:
$(i)$~\textit{Convallaria} data, available online as part of PN2V, consisting of 100 calibration images (diaphragm images, as previously explained) and 100 noisy images of a \textit{Convallaria} section,
$(ii)$~mouse skull nuclei dataset consisting of 500 calibration images (showing the edge of a fluorescent  slide) and 200 noisy realizations of the same static mouse skull nuclei, and
$(iii)$~mouse actin data consisting of 100 calibration images (diaphragm images with only the sample mounting medium in field of view) and 100 noisy realizations of the same static actin sample.
The \textit{Convallaria} and mouse actin datasets are acquired on a spinning disc confocal microscope while the mouse skull nuclei dataset is acquired with a point scanning confocal microscope.

\vspace{1mm}
\noindent \textbf{Implementation and training details:}
All evaluated training schemes are based on the implementation from \cite{krull2019probabilistic} and use the same network architecture: a U-Net~\cite{ronneberger2015u} with depth 3, 1 input channel, and 64 feature channels in the first layer.
All networks are trained with ADAM~\cite{kingma2014adam} with initial learning rate of $0.001$, a patch size of $100$, a batch size of $1$, a virtual batch size of $20$ and the standard learning rate scheduler as used in~\cite{krull2019probabilistic}.
Training is done for $200$ epochs, each consisting of $5$ steps.
We use the \emph{{N2V}} and \emph{{CARE}} (traditional supervised training) implementations from~\cite{krull2019probabilistic}.

With \emph{{PN2V}}, we will refer to the version with the original histogram based noise model, derived form the available calibration data.
As in \cite{krull2019probabilistic}, for each dataset, we use a  $ B \times B$ bin discretization, where $B$ is an integer determined in an empirically optimal manner for which the denoising performance (PSNR) of histogram based PN2V is maximized. 
The minimum and maximum bins are set to the minimum and maximum values present in the  data to be denoised.

Whenever we use our proposed GMM noise model, we will label results with \emph{{PN2V GMM}}. 
As long as not stated differently, all GMM noise models use $K=3$ Gaussians and $n=2$ coefficients per parameter, and are trained on the available calibration data.
Starting from a random initialization, optimization is performed using ADAM with learning rate $0.1$, using a batch size of $25000$ and $4000$ iterations for mouse skull nuclei and mouse actin datasets, and a batch size of $250000$ and $2000$ iterations for the \textit{Convallaria} data.
\sloppy

For bootstrapped PN2V (histogram and GMM based), we use the same setup as for PN2V but naturally taking the bootstrapped noise models instead. They are referred to as {\emph{Boot.~Hist.}} and {\emph{Boot.~GMM}} respectively.
For the latter, we disregard the top and bottom $0.5\%$ percentile of the pseudo GT pixels during noise model training, as we empirically observe that their N2V predictions can be often unreliable.

\fussy
\vspace{1mm}
\noindent \textbf{Comparing different training schemes:} 
For each dataset and denoising method, we repeated each experiment 5 times and then compared the denoised images in terms of peak-signal-to-noise-ratio (PSNR) to available GT images. 
Results can be seen in Fig.~\ref{fig:QualitativeResults}, as well as Table~\ref{QuantitativeTable}. We also evaluated the structural similarity (SSIM) score for all datasets and made them available at {\small \url{github.com/juglab/ppn2v/wiki}}.
\figAblation

Naturally, the fully supervised CARE networks, trained on clean ground truth images, show the best performance on all datasets.
On the mouse actin dataset, PN2V using our GMM based noise model derived from high quality calibration data outperforms all other methods.
Notably, on the other two datasets, our fully unsupervised bootstrapped approach provides superior results and is remarkably close to CARE.
For a discussion of these results, see Section~\ref{sec:discussion}.

\vspace{1mm}
\noindent \textbf{Ablation and parameter study:} Next we compare the robustness of histogram and GMM based noise models with respect to increasingly imperfect calibration data, using the \textit{Convallaria} dataset as an example.
These \textit{ablation studies} consist of two scenarios, where
$(i)$~the available calibration data covers less and less of the range of signals in the data to be denoised, and 
$(ii)$~the amount of available calibration pixels decreases successively.
Figure~\ref{fig:NoiseModelViews}~(c,d) shows example noise models that are derived from ablated calibration data. 
Evidently, for both ablation tests PN2V GMM performance is more robust compared to PN2V (see Fig.~\ref{fig:SystematicTests}).

We also investigated the sensitivity of GMM noise models with respect to the chosen hyper parameters. We performed a parameter study, varying the number of Gaussian kernels $K$ and polynomial coefficients $n$, using the \emph{Convallaria} dataset with full available calibration data.
Results are summarized in Table~\ref{AblationTable}.
While these tests suggest that the simple linear model (one Gaussian, two coefficients) is slightly  preferable, the performance of all configurations remains superior to N2V (see Table~\ref{QuantitativeTable}). 
We additionally measured the performance of a linear noise model using 1 Gaussian and 3 Gaussians with imperfect calibration data (see Table~\ref{gaussiancomparison}). 
We observe that a noise model with 3 Gaussians leads to more stable results.

\section{Discussion}
\label{sec:discussion}
\vspace{-2mm}

\sloppy

We presented a GMM based variation of PN2V noise models and showed that they can achieve higher reconstruction quality even with imperfect calibration data (Fig.~\ref{fig:SystematicTests}).
Additionally, we introduced a novel bootstrapping scheme, which allows PN2V to be trained fully unsupervised using only the data to be denoised (Fig.~\ref{fig:NoiseModelViews}(b)). Our results (Table~\ref{QuantitativeTable}) show that the denoising quality of bootstrapped PN2V is quite close to fully supervised CARE~\cite{weigert2018content} and significantly outperforms N2V~\cite{krull2019noise2void}.
Hence, if calibration data for a given microscope is unavailable, bootstrapping offers an excellent alternative.

\fussy

Interestingly, at times, bootstrapped GMM based noise models even outperform models derived from calibration data.
A possible reason for such good performance is that the distribution of pseudo GT signals used in bootstrapping corresponds well to the distribution of signals in the data to be denoised.
The distribution of GT signals in the calibration data however, can be quite different.

GMM noise models, trained according to Eq.~\ref{eq:opt}, prioritize signals that are abundant in the (pseudo) GT and provide a better fit in these regions compared to others. Figure~\ref{fig:NoiseModelViews}(b) corroborates that our bootstrapped GMM fits well to the true noise distribution for lower signals, which frequently occur in the \textit{Convallaria} data, 
but fails for higher signals.
However, the GMM trained on calibration data (Fig.~\ref{fig:NoiseModelViews}(a)), prioritizes its fit for higher signals, which are frequent in the calibration data, but barely present in the \emph{Convallaria} dataset.

We strongly believe that the methods we propose will help to make high quality DL based denoising an easily applicable tool that does not require the acquisition of paired training data or calibration data.
This would facilitate a plethora of projects in cell biology, where the processes to be imaged are very photosensitive or so dynamic that suitable training image pairs cannot be obtained. 

\section{Acknowledgements}
\label{sec:acknowledgements}
\vspace{-2mm}

The authors thank the LM Facility, Diana Afonso and Jacqueline Tabler from MPI-CBG for kindly sharing samples and expertise, and Matthias Arzt from CSBD/MPI-CBG for helpful discussions. 
Work was supported by the BMBF under the codes 031L0102 (de.NBI) and 01IS18026C (ScaDS2), and the DFG under code JU3110/1-1 (FiSS).

\bibliographystyle{IEEEbib}
\bibliography{strings,refs}

\begin{thebibliography}{10}

\bibitem{weigert2018content}
Martin Weigert, Uwe Schmidt, Tobias Boothe, Andreas M{\"u}ller, Alexandr
  Dibrov, Akanksha Jain, Benjamin Wilhelm, Deborah Schmidt, Coleman Broaddus,
  Si{\^a}n Culley, et~al.,
\newblock ``Content-aware image restoration: pushing the limits of fluorescence
  microscopy,''
\newblock {\em Nature methods}, vol. 15, no. 12, pp. 1090, 2018.

\bibitem{zhang2019poisson}
Yide Zhang, Yinhao Zhu, Evan Nichols, Qingfei Wang, Siyuan Zhang, Cody Smith,
  and Scott Howard,
\newblock ``A poisson-gaussian denoising dataset with real fluorescence
  microscopy images,''
\newblock in {\em CVPR}, 2019.

\bibitem{buchholz2019cryo}
Tim-Oliver Buchholz, Mareike Jordan, Gaia Pigino, and Florian Jug,
\newblock ``Cryo-care: content-aware image restoration for cryo-transmission
  electron microscopy data,''
\newblock in {\em 2019 IEEE 16th International Symposium on Biomedical Imaging
  (ISBI 2019)}. IEEE, 2019, pp. 502--506.

\bibitem{buchholz2019content}
Tim-Oliver Buchholz, Alexander Krull, R{\'e}za Shahidi, Gaia Pigino,
  G{\'a}sp{\'a}r J{\'e}kely, and Florian Jug,
\newblock ``Content-aware image restoration for electron microscopy,''
\newblock {\em Methods Cell Biol}, vol. 152, pp. 277--289, 2019.

\bibitem{krull2019noise2void}
Alexander Krull, Tim-Oliver Buchholz, and Florian Jug,
\newblock ``Noise2void-learning denoising from single noisy images,''
\newblock in {\em Proceedings of the IEEE Conference on Computer Vision and
  Pattern Recognition}, 2019, pp. 2129--2137.

\bibitem{krull2019probabilistic}
Alexander Krull, Tomas Vicar, and Florian Jug,
\newblock ``Probabilistic noise2void: Unsupervised content-aware denoising,''
\newblock {\em arXiv preprint arXiv:1906.00651}, 2019.

\bibitem{belthangady2019applications}
Chinmay Belthangady and Loic~A Royer,
\newblock ``Applications, promises, and pitfalls of deep learning for
  fluorescence image reconstruction,''
\newblock {\em Nature methods}, p.~1, 2019.

\bibitem{sui2018differential}
Liyuan Sui, Silvanus Alt, Martin Weigert, Natalie Dye, Suzanne Eaton, Florian
  Jug, Eugene~W Myers, Frank J{\"u}licher, Guillaume Salbreux, and Christian
  Dahmann,
\newblock ``Differential lateral and basal tension drive folding of drosophila
  wing discs through two distinct mechanisms,''
\newblock {\em Nature communications}, vol. 9, no. 1, pp. 4620, 2018.

\bibitem{laine2019nanoj}
Romain~F Laine, Kalina~L Tosheva, Nils Gustafsson, Robert~DM Gray, Pedro
  Almada, David Albrecht, Gabriel~T Risa, Fredrik Hurtig, Ann-Christin
  Lind{\aa}s, Buzz Baum, et~al.,
\newblock ``Nanoj: a high-performance open-source super-resolution microscopy
  toolbox,''
\newblock {\em Journal of Physics D: Applied Physics}, vol. 52, no. 16, pp.
  163001, 2019.

\bibitem{ouyang2018deep}
Wei Ouyang, Andrey Aristov, Micka{\"e}l Lelek, Xian Hao, and Christophe Zimmer,
\newblock ``Deep learning massively accelerates super-resolution localization
  microscopy,''
\newblock {\em Nature biotechnology}, vol. 36, no. 5, pp. 460, 2018.

\bibitem{noise2noise}
Jaakko Lehtinen, Jacob Munkberg, Jon Hasselgren, Samuli Laine, Tero Karras,
  Miika Aittala, and Timo Aila,
\newblock ``{N}oise2{N}oise: Learning image restoration without clean data,''
\newblock in {\em International Conference on Machine Learning}, 2018.

\bibitem{batson2019noise2self}
Joshua Batson and Loic Royer,
\newblock ``Noise2self: Blind denoising by self-supervision,''
\newblock in {\em International Conference on Machine Learning}, 2019.

\bibitem{ronneberger2015u}
Olaf Ronneberger, Philipp Fischer, and Thomas Brox,
\newblock ``U-net: Convolutional networks for biomedical image segmentation,''
\newblock in {\em MICCAI}, 2015.

\bibitem{kingma2014adam}
Diederik~P Kingma and Jimmy Ba,
\newblock ``Adam: A method for stochastic optimization,''
\newblock {\em arXiv preprint arXiv:1412.6980}, 2014.

\end{thebibliography}

\end{document}